# S&P500 波动率的估计与评价

**摘 要**：在这个衍生品交易全球盛行的时代，风险管理逐渐成为现代金融的核心内容，标的资产的波动率更显得尤为重要。为研究如何准确地估计标准普尔 500 指数的波动率，本文在介绍了各种方法的理论背景以后，采用历史波动率方法、GARCH 模型预测法、隐含波动率法来分别对真实波动率作出估计，同时还考虑了滚动与增加两种调整估计窗的方式，利用无偏性检验与拟合优度检验对几种方法进行了评判。实证的结果说明了隐含波动率是真实波动率的最优估计。在利用历史波动率进行估计时，本文推荐使用滚动估计窗的方式，反之，在利用 GARCH 模型进行预测时用增加估计窗则会表现更佳。

**关键词**：历史波动率；隐含波动率；GARCH 模型

## Volatility of S&P500: Estimation and Evaluation

**Abstract:** In an era when derivatives is getting popular, risk management has gradually become the core content of modern finance. In order to study how to accurately estimate the volatility of the S&P 500 index, after introducing the theoretical background of several methods, this paper uses the historical volatility method, GARCH model method and implied volatility method to estimate the real volatility respectively. At the same time, two ways of adjusting the estimation window, rolling and increasing, are also considered. The unbiased test and goodness of fit test are used to evaluate these methods. The empirical result shows that the implied volatility is the best estimator of the real volatility. The rolling estimation window is recommended when using the historical volatility. On the contrary, the estimation window is supposed to be increased when using the GARCH model.

# 一、引言

## （一）问题的提法

  风险管理是现代金融的核心之一，而波动率（Volatility）则是投资风险的数量体现。风险管理意味着在考虑投资某种资产时，不仅仅要考虑到期望收益，更要考虑到该种资产收益率的波动率，在衍生品交易全球盛行的当下时更是如此。然而未来资产的收益率作为一个分布未知的随机变量，没有人知道其精确的均值与方差，只能通过一些从统计学角度上看似合理的方式利用历史收益率作出参数估计与假设检验。对于均值的估计与检验已不罕见，而对于方差（即波动率）的讨论则尚在完善，本文的研究重点便是以标准普尔 500（SPX）指数为例，根据 2004-2019 年的历史收益率数据，并利用假设检验的方法评判几种波动率估计方法的优劣。

## （二）研究方法概述

### 1.研究主题

我们将使用不同的估计方法来估计 SPX 的真实波动率（Realized Volatility），并通过一些标准对它们进行比较。在比较时，我们将使用年化波动率，但估计窗（Estimation Window）将每月更改，也意味着我们将每月计算一次波动率。

### 2.数据说明

(1)SPX 日收益率序列：2004 年-2019 年，日度数据；

(2)VIX 日收盘价：2005 年-2019 年，日度数据。

数据均来源于 Choice 金融终端。

### 3.估计方法

(1)基于历史数据的样本方差（Historical Sample Variance）：简单计算估计窗的样本方差（无偏），窗口的更迭方式可以选用滚动（Rolling）与增加（Increasing）两种；

(2)基于历史数据的广义自回归条件异方差模型（Historical GARCH）：通过 ARCH 检验判断收益率序列是否具有异方差性（Heteroscedasticity），如果有，则可以用 GARCH 模型对其进行方差预测，同样，窗口的更迭方式可以选用滚动与增加两种；

(3)隐含波动率：利用隐含波动率来对真实波动率进行预测，在这之前本文会介绍 Black-Scholes、Model-Free 和 Corridor 隐含波动率（Implied Volatility）的计算方法，实证中会直接利用芝加哥期权交易所（CBOE）推出的 VIX 指数，它就是一个跟踪 SPX 波动率的指数。

### 4.评价方法

(1)无偏性检验（Unbiased Test）：通过建立一个线性回归模型来探讨各个估计方法的"实证无偏性"，这种无偏性是建立在一个"截距项为 0、系数项为 1"的原假设下的，通过适当变形并比较 $F$ 统计量与 $p$ 值的大小可以大致评判几种估计方法的优良性；

(2)拟合优度检验（Goodness of Fit Test）：在均方误差（Mean Square Error）的视角下对几种方法进行比较，并在正态假设下对原假设"估计波动率的期望为真实波动率"进行检验，通过比较 $\chi^2$ 统计量与 $p$ 值的大小可以大致评判几种估计方法的优良性。

### 5.符号说明

由于我们将使用多种方法来对波动率进行估计，在正式开始探讨之前，要对符合进行注释，以防止产生歧义。

$r_t$:第 $t$ 期 SPX 的收益率；

$\sigma_t$:第 $t$ 期 SPX 的真实波动率；

$\hat{\sigma}_t$:第 $t$ 期 SPX 的估计波动率；

$T$:估计窗的长度。

## 二、历史波动率

### （一）样本方差估计法

当我们处于 2020 年 1 月 1 日，我们很难去想象 SPX 在 2020 年 1 月的真实波动率是多少，但是我们可以通过回顾 2019 年全年 SPX 的行情，来对 2020 年 1 月作出一定窥探。

根据统计学的基本知识，样本方差 $\hat{\sigma}^2 = \frac{1}{n-1}\sum_{t=1}^{n}(r_t - \bar{r})^2$ 是总体方差 $\sigma^2$ 的无偏估计，其中 $\bar{r} = \frac{1}{n}\sum_{t=1}^{n} r_t$ 是样本均值。无偏估计纵然很好，但是很可惜，总体方差 $\sigma^2$ 并不一定等于未来的真实波动率，所以历史数据构造的样本方差并一定不是未来波动率的无偏估计。

然而，我们知道所有资产的期望收益率都有均值回复（Mean Reversion）现象，实际上长期而言，波动率也不例外。在大样本情形下，人们总认为过去的样本信息可以作为未来的估计，并且在构建资产组合或者进行收益预测的时候也一贯这么去做。虽然在很多情形下，它们并不合理且也没有获取正向的异常收益，但是由于其操作简单以及思路的清晰性，仍然广受追捧。

在本文中，我们也将样本方差估计法作为一种方法，来对真实收益率作出估计，并与其他估计方法进行一系列比较，得出方法的优劣性以及操作时的一些推荐。

为使得一切估计波动率和真实波动率可以进行比较，我们将所有波动率换算为年化波动率，考虑到我们收集到的数据是 SPX 的日度收益率，故用

$$\hat{\sigma}_t^2 = \frac{252}{T}\sum_{i=t-T}^{t-1}(r_i - \bar{r})^2 \tag{2-1}$$

来表示利用长度为 $T$ 的估计窗来对 $\sigma_t^2$ 作出的估计，252 是美国证券市场一年的开盘日数，也可以说是金融工作者一年的工作时长。

在估计中，我们选用滚动估计窗与增加估计窗两种方法，滚动估计窗方法意味着在作每一次估计时，总利用该日前一年的数据来进行估计；而增加估计窗方法意味着作每一次估计时，总利用该日以前所有的数据来进行估计。简而言之，滚动估计窗意味着 $T$ 恒定，而增加估计窗意味着 $T$ 是 $t$ 的线性递增函数。

本文认为，就历史样本方差估计法而言，滚动估计窗的方法应该会比增加估计窗的方法表现更出色。原因是：滚动估计窗仅仅考虑近期的数据，近期数据的趋势以及属性会更类似我们要估计的时段，更能体现待估时段的特性；增加估计窗则考虑所有历史数据，特性在时间更迭中慢慢被磨灭，这种方法更注重共性。共性固然重要，但对于估计未来一个短期波动率而言，特性或许才更能说明问题。

### （二）GARCH 模型估计法

历史方差估计法在隐含波动率问世之后就已经逐渐没落，但计量经济学的发展（主要是 ARCH 族、GARCH 族模型的问世）又使得历史波动率重新回到人们视野中。Andersen 和 Bollershev

（1998）[1]通过实证证实了 GARCH 模型对于波动率拥有非常好的预测能力，这也启示了我们必须将 GARCH 模型也加入讨论中来。

经典的线性回归或自回归探讨的是方差齐性的时间序列的均值问题，但 SPX 收益率显然具有异方差性（否则我们也就不必要探讨这个话题了），通过 GARCH 模型对其方差进行建模的方法也早已流行起来。本文也通过滚动估计窗与增加估计窗两种方法来利用 GARCH 模型对未来波动率进行预测，并评价其预测能力。

我们还是首先绘制出收益率序列的图像，来观察其是否具有异方差性，如图 1 所示。

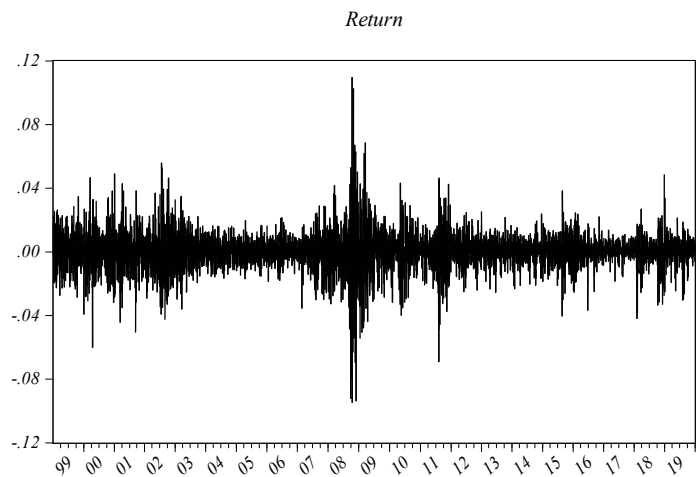

图 1 SPX 收益率走势图

很容易看出 SPX 的收益率是具有异方差性的，最明显的是在 2003 年开始到 2007 年，SPX 的波动率非常小，说明美国股票市场这段时间内保持一种非常稳定的状态。但天有不测风云，2008 年金融危机来袭，美国作为金融危机发源国，其标志性金融市场指数 SPX 的风险也逐渐增大，这段时间内的收益率波动性远远超过其他时期。2010 年以后的图像整体看似也参差不齐但似乎也有周期性、规律性可循，但无论如何，根据图像来看，SPX 指数收益率拥有较严重的异方差性。

虽然我们知道收益率序列必定是平稳序列，但在对时间序列展开研究的过程中，依然要对数据进行单位根检验。我们选用 ADF 方法，ADF 检验的统计量是-56.07154，$p$ 值则是 0.0001，这足以说明收益率序列是平稳过程。

其次，在提取 SPX 收益率的均值后，我们对序列进行 ARCH 检验，以真正证明其具有 ARCH 效应，ARCH 检验的检验结果如下所示：

| $F$-统计量 | 242.7837 | Prob.$F$ | 0.0000 |
| $\chi^2$ 统计量 | 232.2179 | Prob.$\chi^2$ | 0.0000 |

很显然，无论是 $F$ 统计量或是 $\chi^2$ 统计量，都能够在 0.01 的显著性水平下拒绝原假设，说明 SPX 收益率序列存在严重的异方差性。不过，在我们建立 GARCH(2,1)模型以后，再对残差进行 ARCH 检验，发现 ARCH 效应已经消失。由此受到启发，我们将在波动率的估计中采取 GARCH(2,1)模型进行预测。但是考虑到 GARCH 模型在作出预测时候有可能出现的过度拟合（也就是过度考虑增长趋势的现象），我们需要在 GARCH 模型的预测波动率上作一个合理的限制：如果估计的

波动率比上一期真实波动率的 2 倍还大，那么投资者往往不会相信 GARCH 模型估计的波动率是具有优良性质的，故修正后的估计波动率为

$$\hat{\sigma} = \min\{GARCH\text{估计的波动率},\text{上期波动率的}2\text{倍}\}. \tag{2-2}$$

当然，上文已经提及，要通过滚动估计窗与增加估计窗两种方法来利用 GARCH 模型对未来波动率进行预测。本文认为，与历史样本方差估计法不同，在利用 GARCH 模型进行波动率预测时，增加估计窗或许将比滚动估计窗表现得更好。原因在于：虽然在历史样本方差估计法中提到过，久远的数据可能太旧了以至于对未来的估计没有作用，但 GARCH 模型与历史样本方差估计法不同，GARCH 模型考虑到了趋势，它已经不单单像历史样本方差估计法中那样简单地求一个平均值了，久远的数据对于我们研究趋势的规律仍然具有帮助，它们有助于降低估计系数的标准误差，对于预测而言也更加准确。

## 三、隐含波动率

### （一）Black-Scholes 公式与 Vega

Black-Scholes 期权定价公式的提出被誉为华尔街的第二次革命，借助伊藤引理（Ito's Lemma）与热传导方程的帮助，Black 和 Scholes（1973）[2]给出了期权价格的解析表达式，尽管其表达式是由标准正态分布的分布函数所呈现，也不免令人惊叹金融中蕴藏的神奇奥秘。除了用以期权定价，Black-Scholes 公式还被广泛地运用于其他领域，例如将市场上期权的买卖价格回代入 Black-Scholes 公式求解出的波动率称为隐含波动率。Black-Scholes 公式直到现今仍然是衍生品交易中最具影响力的公式。

当然金融工程理论发展到现在，已经可以利用更简单的例如二叉树、鞅（Martingale）等方法求解出 Black-Scholes 公式。设标的资产的当前价格是 $S_0$，无风险利率为 $r$，期权的行权价为 $X$，假设到期日 $T$ 时资产价格有 $\ln S_T \sim N\left(\ln S_0 + \left(r - \frac{\sigma^2}{2}\right)T, \sigma^2 T\right)$，则欧式看涨期权的价格可以表示为

$$C = e^{-rT}E^Q[S_T - X]^+, \tag{3-1}$$

其中 $E^Q(\cdot)$ 是风险中性期望，$e^{-rT}E^Q(\cdot)$ 表示的是未来期权收益在 0 时刻的现值，$[\cdot]^+$ 则表示取正部。对该期望作积分变换以后可以解出看涨期权的价格

$$C = S_0 \Phi(d_1) - Xe^{-rT}\Phi(d_2), \tag{3-2}$$

其中，$d_1 = \dfrac{\ln\left(\frac{S_0}{X}\right) + \left(r + \frac{1}{2}\sigma^2\right)T}{\sigma\sqrt{T}}$，$d_2 = \dfrac{\ln\left(\frac{S_0}{X}\right) + \left(r - \frac{1}{2}\sigma^2\right)T}{\sigma\sqrt{T}} = d_1 - \sigma\sqrt{T}$。而利用期权平价公式 $C + Xe^{-rT} = P + S_0$ 也可以求出同一行权价格的看跌期权价格 $P$。

通过 Black-Scholes 公式我们知道，如果知道标的资产的价格、无风险利率、行权价格、标的资产波动率与到期日，就可以很容易的求出期权的价格。但在现实市场中，波动率往往是难以衡量的，期权价格倒是可以通过市场买卖双方均衡的机制而自主生成。这时我们将期权价格代入到

Black-Scholes 公式反解出波动率，以这种方式求得的波动率称为隐含波动率。

定义 $Vega = \frac{\partial C}{\partial \sigma} = S\sqrt{T}\varphi(d_1)$，$Vega$ 代表的是期权价格对标的资产波动率的敏感程度，很明显当 $d_1 = 0$ 时，$Vega$ 最大，即行权价格为 $X_0 = Se^{\left(r+\frac{1}{2}\sigma^2\right)T}$ 的期权关于标的资产波动率的敏感程度最大，这一期权又被称为最大 $Vega$ 期权。我们定义 $Vega$ 值的意图不仅仅是用来探究期权价格对于波动率的敏感程度，而且还考虑到了市场上有各种各样行权价格的期权，用不同的期权来计算隐含波动率会出现不同的结果（著名的波动率微笑），而最大 $Vega$ 期权是对资产波动率的敏感程度最大的期权，选择最大 $Vega$ 期权来计算隐含波动率会得到相对准确的结果。

### （二）Model-Free 隐含波动率

Black-Scholes 公式的横空出世使得衍生品交易成为全球焦点，但是资产价格服从对数正态分布的要求实在是太过苛刻，这在现实世界的很多情况下是不成立的。无模型（Model-Free）隐含波动率受到 Black-Scholes 公式的启发，但不对资产价格作对数正态分布的假设来求解隐含波动率。

Model-Free 隐含波动率起源于方差互换（Variance Swap），最早出现对方差互换的研究是 Dupire（1994）[3] 和 Neuberger（1994）[4] 的文章，基于对方差互换的研究，Demeterfi，Derman，Kamal 和 Zou（1999）[5] 推导出了远期方差的公平价值。Britten 和 Neuberger（2000）[6] 则更进一步，在比 Black-Scholes 模型更简单的假设下，推导出了 Model-Free 隐含波动率。

在我国，对 Model-Free 隐含波动率展开的研究很少，国外文献也一贯使用复杂的证明方式来进行推导，本文在总结国内外文献的基础上，以较为浅显易懂的方式给出 Model-Free 隐含波动率的推导过程，其核心思想是方差互换。

假设资产价格服从随机微分方程 $\frac{dS_t}{S_t} = rdt + \sigma_t dW_t^Q$，其中 $W_t^Q$ 是风险中性测度下的标准维纳过程，我们可以在这个随机微分方程里看出它与 Black-Scholes 假设下不一样的地方，即其方差项是时变的，这可以说是一个非常宽泛的模型假设了。

方差互换合约的含义是：在 $T$ 时刻，合约卖方向买方支付

$$\frac{1}{n\Delta}\sum_{k=1}^{n}\left(\frac{S_{t_k}-S_{t_{k-1}}}{S_{t_{k-1}}}\right)^2, \tag{3-3}$$

其中 $t_k = k\Delta$，且 $n\Delta = T$. 为公平，买方应在 0 时刻支付一笔现金给卖方。可以发现，式(3-3)很像是描绘了 $[0,T]$ 时间段上股票的波动率，事实上如果 $\Delta$ 很小，有

$$\sum_{k=1}^{n}\left(\frac{S_{t_k}-S_{t_{k-1}}}{S_{t_{k-1}}}\right)^2 \approx \int_0^T \left(\frac{dS}{S}\right)^2 = \int_0^T \sigma_t^2 dt, \tag{3-4}$$

如果我们要预测 $[0,T]$ 时间段上的波动率 $\frac{1}{T}\int_0^T \sigma_t^2 dt$，那么对方差互换合约作出合理的定价实际上就已经是对波动率作出了良好预测。如果 $\Delta$ 很小，近似有

$$\ln S_{t_k} - \ln S_{t_{k-1}} = \ln\left(1 + \frac{S_{t_k} - S_{t_{k-1}}}{S_{t_k}}\right) = \frac{S_{t_k} - S_{t_{k-1}}}{S_{t_{k-1}}} - \frac{1}{2}\left(\frac{S_{t_k} - S_{t_{k-1}}}{S_{t_{k-1}}}\right)^2, \tag{3-5}$$

上式两侧求和，得到

$$\ln S_T - \ln S_0 = \sum_{k=1}^{n} \frac{S_{t_k} - S_{t_{k-1}}}{S_{t_{k-1}}} - \frac{1}{2}\sum_{k=1}^{n}\left(\frac{S_{t_k} - S_{t_{k-1}}}{S_{t_{k-1}}}\right)^2, \tag{3-6}$$

即

$$\sum_{k=1}^{n}\left(\frac{S_{t_k} - S_{t_{k-1}}}{S_{t_{k-1}}}\right)^2 = 2\ln\frac{S_0}{S_N} + 2\int_0^T \frac{dS}{S}$$
$$= 2\ln\frac{S_0}{S_N} + 2rT + \int_0^T \sigma_t dW_t^Q, \tag{3-7}$$

两侧在 $\mathcal{F}_0$ 下求期望，由于关于布朗运动的随机积分的鞅性，有

$$E^Q\left[\int_0^T \sigma_t dW_t^Q\right] = 0, \tag{3-7}$$

剩下只需计算 $\ln\frac{S_0}{S_T}$ 的期望，易证

$$\ln\frac{S_0}{S_T} = \ln\frac{S_0}{F_0} + \ln\frac{F_0}{S_T}$$
$$= -rT + \frac{F_0 - S_T}{F_0} + \int_0^{F_0} \frac{\max\{S_T - x, 0\}}{x^2}dx + \int_{F_0}^{+\infty} \frac{\max\{x - S_T, 0\}}{x^2}dx, \tag{3-8}$$

其中 $F_0 = e^{rT}S_0$ 是标的资产对应的期货合约，注意到

$$E^Q\left[\frac{F_0 - S_T}{F_0}\right] = 0 \tag{3-9}$$

以及

$$E^Q\left[\int_0^{F_0} \frac{\max\{S_T - x, 0\}}{x^2}dx + \int_{F_0}^{+\infty} \frac{\max\{x - S_T, 0\}}{x^2}dx\right] = e^{rT}\int_0^{+\infty} \frac{\max\{c(x), p(x)\}}{x^2}dx, \tag{3-10}$$

其中 $c(x), p(x)$ 分别表示行权价格为 $x$ 的看涨和看跌期权，故最终得到方差互换的定价结果是

$$e^{-rT}E^Q\left[\frac{1}{n\Delta}\sum_{k=1}^{n}\left(\frac{S_{t_k} - S_{t_{k-1}}}{S_{t_{k-1}}}\right)^2\right] = \frac{2}{T}\int_0^{+\infty} \frac{\max\{c(x), p(x)\}}{x^2}dx, \tag{3-11}$$

可以看出，等式左侧就是方差互换合约的定价，而等式右侧则是一个将市场中所有的看涨、看跌期权囊括在一起构造的无穷积分，这就是 Model-Free 隐含波动率。

### （三）Corridor 隐含波动率与 VIX 指数

Black-Scholes 公式只考虑一种行权价格的期权（通常是最大 *Vega* 期权），而根据 Model-Free 隐含波动率的解析表达式可以看出，Model-Free 隐含波动率考虑市场上所有行权价格的期权。前者考虑过少，特别是期权的选取会导致结果产生比较大的误差；后者考虑过多，在真实市场中，行权价格过高或过低的期权往往不会受到投资者亲睐。

Corridor 隐含波动率的推导较为复杂，但其本质上就是在 Model-Free 隐含波动率的基础上将

积分区域限制在一个有限区间中，很符合其走廊（Corridor）的名头。在我国，对于 Corridor 隐含波动率的研究较少，而芝加哥期权交易市场早在 2003 年就已经利用 Corridor 隐含波动率来编制 VIX 指数，跟踪 SPX，时至今日，VIX 指数已经不仅成为了全球关注的"恐慌指数"，更成为了全球风险管理的风向标。

Corridor 隐含波动率的表达形式很简单，即

$$\hat{\sigma}^2 = \frac{2}{T}\int_L^U \frac{\min\{P(x),C(x)\}}{x^2}dx \tag{3-12}$$

它表达的含义是抛开行权价格过高或者行权价格过低的期权，只考虑市场中真正发生流通的那些期权。通过这些期权来计算隐含波动率，最终结果将比 Model-Free 隐含波动率更加准确，现实中的 $U$ 和 $L$ 常常取上下 25%分位数或 10%分位数。值得注意的是，将计算出来的波动率乘上 100，就可以近似地编织出 VIX 指数。

## 四、估计波动率的评价方法

### （一）线性回归与实证无偏

基于下述线性回归模型

$$\sigma_t = a + b\hat{\sigma}_t + \varepsilon_t, \ \varepsilon_t \sim N(0,\delta^2) \tag{4-1}$$

我们探讨一个假设检验问题：

$$H_0: a=0, b=1 \text{ vs } H_1: \text{Complement of } H_0。$$

可以试想，如果原假设成立，则有 $\sigma_t = \hat{\sigma}_t + \varepsilon_t$，这也就意味着可以将 $\hat{\sigma}_t$ 视作 $\sigma_t$ 的"无偏估计"，当然这个无偏估计是一个实证无偏估计，是根据实证结果得到的"无偏性"。

对于上述假设检验问题，很自然的想法是通过似然比检验法构造拒绝域，根据方差分析中的平方和分解原理将检验统计量的分子分母化为相互独立的 $\chi^2$ 分布，最后简化为一个 $F$ 统计量。但是可以运用更巧妙的思路，即在回归方程左右同时减去 $\hat{\sigma}_t$，得到：

$$\sigma_t - \hat{\sigma}_t = a + b\hat{\sigma}_t + \varepsilon_t, \ \ \varepsilon_t \sim N(0,\delta^2) \tag{4-2}$$

当然注意这里的 $b$ 和原来的 $b$ 不是同一个，此时发现最初的假设检验问题等价于该模型下的另一个假设检验问题：

$$H_0: a=0, b=0 \text{ vs } H_1: \text{Complement of } H_0$$

这样的假设检验问题就成为了经典线性回归中检验模型整体显著性的 $F$ 检验，在任何统计软件中都可以直接导出它的统计量与 $p$ 值。

当然，想在这样一个"苛刻"的检验下接受原假设是很难的，毕竟在 2005-2019 年这样一个大样本下两个时间序列总有一些统计规律可循。因此，去比较哪些估计方法接受了原假设不太现实，更好的思路是比较 $F$ 统计量，如果 $F$ 统计量更低则说明该估计方法更好。再退一步说，可以去参考截距项与系数项的回归结果与 $p$ 值来分析问题，追求的结果自然是不显著或者显著为一个绝对

值很小的常数。

## （二）均方误差与拟合优度检验

比较均方误差与拟合优度是衡量估计误差程度常用的方法，我们首先对均方误差进行介绍，然后再利用均方误差来推导此时的拟合优度检验。

均方误差，顾名思义，就是平方误差的平均值，其计算公式为

$$MSE = \frac{1}{N}\sum_{t=1}^{N}\left(\hat{\sigma}_t - \sigma_t\right)^2 \tag{4-3}$$

可以看出，均方误差越小，则说明该方法的估计精度是越准确的，在数理统计学中，均方误差最小的估计又称为有效估计，故均方误差的比较又称为有效性比较，而如果某个估计在无偏估计中最有效，则该估计称为最小方差无偏估计（UMVUE）。本文在后续就将利用均方误差的思想来比较几种估计波动率的方法。

拟合优度说明的也是估计量与待估参数的接近程度，当然，本文利用均方误差构造的拟合优度检验是建立在一个正态假设下的。

假设 $\hat{\sigma}_t \sim N(\sigma_0, \delta^2)$，我们去考虑假设检验问题：

$$H_0: \sigma_0 = \sigma \quad vs \quad H_1: \sigma_0 \neq \sigma$$

我们不用传统的 $t$ 检验去探讨该问题，因为那样意义不大，相当于只是单纯地求了一次平均值，我们要在均方误差的思想下来进行拓展，延伸出均方误差对应的检验 $p$ 值。

显然，在原假设为真的前提下，$\frac{\hat{\sigma}_t - \sigma}{\delta} \sim N(0,1)$，根据正态分布与卡方分布的关系又能够得到 $\left(\frac{\hat{\sigma}_t - \sigma}{\delta}\right)^2 \sim \chi^2(1)$，而根据 $\chi^2$ 分布的可加性与均方误差的定义，我们得到

$$\frac{N \cdot MSE}{\delta^2} \sim \chi^2(N) \tag{4-4}$$

不过由于 $\delta$ 是未知的，我们仍然不能构造出严格或者近似的拒绝域。但注意到样本方差 $\hat{\delta}^2 = \frac{1}{N}\sum_{t=1}^{N}\left(\hat{\sigma}_t - \sigma\right)^2$ 是 $\delta^2$ 的有效估计，同时 $\hat{\delta}^2 \xrightarrow{a.s.} \delta^2$，即 $\frac{\hat{\delta}^2}{\delta^2} \xrightarrow{a.s.} 1$，根据 Slutsky 定理，有

$$\frac{N \cdot MSE}{\hat{\delta}^2} = \frac{N \cdot MSE}{\delta^2} \cdot \frac{\delta^2}{\hat{\delta}^2} \xrightarrow{d} \frac{N \cdot MSE}{\delta^2} \sim \chi^2(N) \tag{4-5}$$

这也就意味着 $\frac{N \cdot MSE}{\hat{\delta}^2} \sim \chi^2(N)$ 在 $N$ 相当大时近似成立，可以通过该 $\chi^2$ 统计量来推导拒绝域并计算对应的 $p$ 值。由于该方法借助了均方误差且通过 $\chi^2$ 统计量衡量了估计接近程度，所以称其为拟合优度检验。

# 五、实证分析与结果

## （一）实证估计结果

当我们滚动或者增加估计窗时，我们可以每月得到一个真实波动率以及五种估计波动率（四个历史波动率和一个隐含波动率），这五种波动率的输出结果如下表所示。

表 2 波动率的输出结果

| Month | Realized | Historical Sample Variance | | Historical GARCH | | VIX |
|---|---|---|---|---|---|---|
| | | Rolling | Increasing | Rolling | Increasing | |
| 2005.1 | 10.38% | 11.09% | 11.09% | 11.07% | 11.07% | 14.08% |
| 2005.2 | 10.29% | 11.03% | 11.04% | 11.01% | 11.02% | 12.03% |
| 2005.3 | 10.03% | 11.11% | 10.99% | 11.09% | 10.97% | 12.50% |
| 2005.4 | 14.36% | 10.62% | 10.95% | 10.60% | 10.93% | 14.02% |
| ............ | | | | | | |
| 2019.9 | 12.36% | 17.44% | 18.39% | 59.08% | 27.11% | 19.07% |
| 2019.10 | 9.71% | 17.63% | 18.36% | 24.71% | 24.71% | 15.43% |
| 2019.11 | 10.46% | 16.29% | 18.33% | 19.43% | 19.43% | 13.41% |
| 2019.12 | 8.93% | 15.61% | 18.30% | 16.08% | 20.92% | 13.30% |

根据表 2 的结果可以看出，从早期来看（2005.1-2005.3），历史波动率的估计效果明显比隐含波动率要好，这是因为这三个月间 SPX 的波动率整体而言保持稳定，所以用历史数据估计会更加准确，在这种稳定状态下，隐含波动率的估计则明显高估了真实波动率。然而在 2005 年 4 月，真实波动率发生了上升，历史波动率的估计准确程度就开始大大下降，隐含波动率在此时占据很大优势。

从后期来看（2019.9-2019.12），我们发现了用增加估计窗的历史样本方差估计法中的估计波动率已经趋于平稳，这显然不符合短期段波动率的波动方式。滚动估计窗方法下的 GARCH 模型有我们所说的"过度趋势"的情况（在 2019.9，其估计的波动率为 59.08%）。滚动估计窗方法下的历史样本方差估计、增加估计窗方法下的 GARCH 模型估计和隐含波动率是三种更好的估计方法。

当然，通过一个表格很难详尽地看出几种估计方法的优劣程度，我们绘制出各种估计方法以及真实波动率的趋势图，以窥究竟。

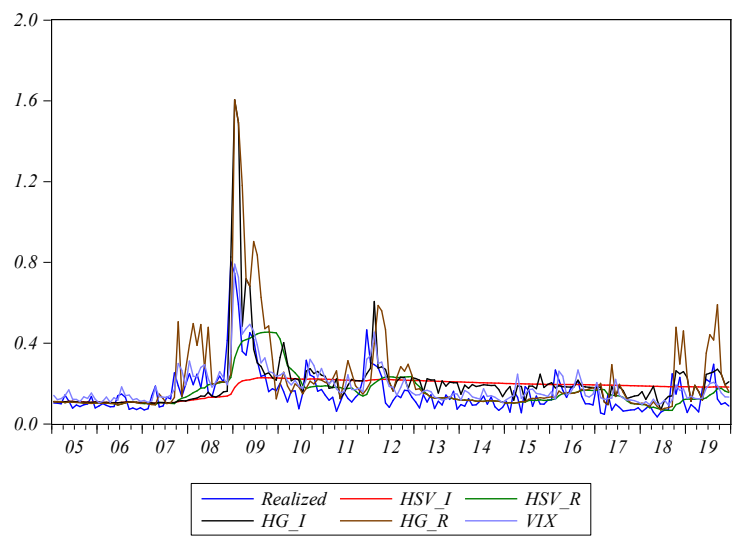

**图 2 波动率走势图**

如图所示，VIX 显然是真实波动性的最佳估计量，它的曲线走势与真实波动率的走势是最为相似的。关于 GARCH 模型，虽然这种方法可以消除异方差性，但它们对波动性上升的反应太过敏锐，特别是 2009 年，GARCH 模型作出了很多不合常理的估计。然而，增加估计窗可以稍微缓和 GARCH 模型出现的这个问题。在历史样本方差估计法中，滚动估计窗的表现似乎优于增加估计窗，从图中看出，增加估计窗得到的估计波动率在后期趋于平稳，这显然不符合短期段波动率的走势。

## （二）检验与评价

现在我们通过上文提到的两个检验法来对五种估计波动率进行量化评估。

### 1.无偏性检验

基于线性回归模型

$$\sigma_t - \hat{\sigma}_t = a + b\hat{\sigma}_t + \varepsilon_t, \quad \varepsilon_t \sim N(0,\delta^2) \tag{5-1}$$

我们考虑假设检验问题：

$$H_0: a=0, b=0 \text{ vs } H_1: \text{Complement of } H_0$$

很容易看出，该检验即为线性回归模型中用以探讨模型整体显著性的 $F$ 检验，在任何统计软件中都可以导出它的统计量与 $p$ 值。在此，我们对五种估计波动率向真实波动率作出回归模型，并探讨其统计显著性，以对它们的优良性作出一定的判断。

**表 3 回归参数与显著性**

| Methods | Historical Sample Variance | | Historical GARCH | | VIX |
|---|---|---|---|---|---|
| | Rolling | Increasing | Rolling | Increasing | |
| $a$ | 0.0800 | 0.0598 | 0.0530 | 0.1084 | -0.0146 |
| *Prob.(a)* | 0.0000 | 0.0000 | 0.0005 | 0.0022 | 0.1476 |
| $b$ | -0.6672 | -0.5447 | -0.3853 | -0.7493 | -0.1009 |
| *Prob.(b)* | 0.0000 | 0.0000 | 0.0000 | 0.0001 | 0.0350 |
| $F$ | 574.1274 | 307.5958 | 22.8440 | 15.7506 | 4.5122 |
| *Prob.(F)* | 0.0000 | 0.0000 | 0.0000 | 0.0001 | 0.0350 |

根据表 3 的结果，我们发现没有任何一个估计波动率是真实波动率的"无偏估计"，当然这并不可惜，因为在样本量如此大的情形下，我们很容易得到非常显著的统计关系，但它仍然具备帮助我们分析问题的资格。

从我们探讨的假设检验问题的角度来看，如果说 $F$ 统计量越小，则说明该估计方法越能代表真实波动率的"无偏估计"。因此，一目了然，表 3 从右往左排列便正好是波动率估计优良性的排名，隐含波动率仍然是最佳的估计，GARCH 模型的估计次之，历史样本方差估计法则在 $F$ 统计量这一角度下远远落后于其他估计方法。

我们探讨截距项与系数项们的估计值与其显著性，我们发现隐含波动率仍是最优的，其截距项不显著，而系数项则是五者之中绝对值最小的，这足以说明在无偏检验视角下，隐含波动率的优良性远远在其他估计方法之上。GARCH 模型的 $F$ 统计量也不大，至少和历史样本方差估计法比起来少了一个数量级，因此在 GARCH 模型和历史样本方差估计法中，GARCH 模型是占优的。

受到郑振龙和黄薏舟（2010）[16]的启发，他们运用多元线性回归与 Wald 检验来探讨隐含波动率是否包含了 GARCH 模型预测波动率的所有信息，思路是：如果将隐含波动率、GARCH 预测波动率一起作为自变量进行线性回归后，GARCH 预测波动率的系数变为不显著且 Wald 检验接受原假设则认为隐含波动率包含了 GARCH 模型预测波动率的所有信息。

在本文的探讨中，由于发现了 Wald 检验的效果不明显（在郑振龙和黄薏舟的文章中也有出现该现象），故我们仍然利用似然比检验。

我们选用在历史样本方差估计法和 GARCH 模型预测法中表现较好的两项，与 VIX 预测波动率一起作为解释变量进行回归，例如我们构造模型：

$$\sigma = a + b_{HSV}\sigma_{HSV} + b_{VIX}\sigma_{VIX} + \varepsilon \tag{5-2}$$

然后去探讨各个系数的估计值与假设检验问题

$$H_0: a=0, b_{HSV}=0, b_{VIX}=1 \text{ vs } H_1: \text{Complement of } H_0$$

的 $F$ 统计量与 $p$ 值，我们依然采用广义似然比检验而并非郑振龙等所用的 Wald 检验。此外特别要注意的是，由于 GARCH 模型被普遍认为优于历史方差估计法，故在该二者中，我们去探讨 GARCH 预测是否包含了历史方差估计法的所有信息。最终的回归与检验统计量如表 4 所示。

表 4 多变量回归结果

|  | $a$ | $HSV_R$ | $HG_I$ | VIX | Adj-$R^2$ | D-W | F |
|---|---|---|---|---|---|---|---|
| *Coefficient* | 0.0392 | 0.1876 | 0.4067 |  | 0.5588 | 1.1524 | 162.1681 |
| *p* | 0.0008 | 0.0101 | 0.0000 |  |  |  | 0.0000 |
| *Coefficient* | -0.0077 | -0.1362 |  | 0.9813 | 0.6710 | 1.5100 | 4.0590 |
| *p* | 0.4715 | 0.0615 |  | 0.0000 |  |  | 0.0189 |
| *Coefficient* | -0.0059 |  | 0.1226 | 0.7176 | 0.6771 | 1.4810 | 5.7830 |
| *p* | 0.5732 |  | 0.0094 | 0.0000 |  |  | 0.0037 |
| *Coefficient* | -0.0004 | -0.1186 | 0.1150 | 0.8006 | 0.6802 | 1.5495 | 4.8054 |
| *p* | 0.9716 | 0.0998 | 0.0147 | 0.0000 |  |  | 0.0030 |

根据表 4 的结果显示：

GARCH 模型预测波动率包含了历史样本方差估计法的部分信息（历史样本方差法在 0.01 的水平下不显著了），但是很难说 GARCH 模型预测波动率包含了其所有信息，因为似然比检验无法接受原假设且 GARCH 模型对应的系数与 1 相去甚远。

第二个模型中，常数项与历史方差对应的波动率都不显著，而且给的显著性水平 0.01，似然比检验接受原假设，这充分说明了隐含波动率的确包含了历史样本方差估计法的所有信息。

在将 GARCH 模型预测波动率与隐含波动率一起回归时，除了常数项不显著以外，包括似然比检验在内的统计量都无法落于接受域，但 GARCH 预测波动率的 $p$ 值变大了，这说明隐含波动率包含了部分 GARCH 模型预测波动率的信息。

在将三个波动率都纳入考虑范围时，结果基本一致，很明显看出历史样本方差估计法是可有可无的，而隐含波动率则是最为显著的，但似然比检验拒绝了原假设，说明隐含波动率包括了另外两个预测波动率一定的信息。

综上所述，我们发现隐含波动率包含了历史方差估计波动率的所有信息、包含了 GARCH 模型预测波动率的部分信息，GARCH 模型预测波动率也包含了历史方差估计波动率的部分信息。

**2.均方误差与拟合优度检验**

假设 $\hat{\sigma}_t \sim N(\sigma_0, \delta^2)$，我们去考虑假设检验问题：

$$H_0: \sigma_0 = \sigma \ \text{vs} \ H_1: \sigma_0 \neq \sigma$$

我们在上文已经提及，在原假设成立的前提下，借助 Slutsky 定理的帮助，我们得到了 $\frac{N \cdot MSE}{\hat{\delta}^2} \xrightarrow{d} \chi^2(N)$，这意味着我们可以借助均方误差展开检验并得到 $p$ 值。

通过计算均方误差以及估计 $\hat{\sigma}_t$ 的方差，我们可以比较出各种估计方法的优良性，比较结果如表 5 所示。

表 5 均方误差与拟合优度检验

| Methods | Historical Sample Variance | | Historical GARCH | | VIX |
|---|---|---|---|---|---|
| | Rolling | Increasing | Rolling | Increasing | |
| $MSE$ | 0.0099 | 0.0131 | 0.0316 | 0.0170 | 0.0050 |
| $\hat{\delta}^2$ | 0.0076 | 0.0018 | 0.0466 | 0.0306 | 0.0096 |
| $p$ | 0.0032 | 0.0000 | 0.9997 | 1.0000 | 1.0000 |

根据表 5 的结果，我们可以看出，在均方误差的标准下，隐含波动率仍然是真实波动率的最佳估计，历史样本方差方法的均方误差也不大，GARCH 模型估计出来的波动率具有较大的均方误差。针对两种历史波动率估计方法而言，滚动估计窗与增加估计窗的优劣与最初分析的是一致的，对于历史样本方差方法，滚动估计窗可以作出更有效的估计，而对于 GARCH 模型方法，纳入更多的数据进入样本可以增强 GARCH 模型参数估计的精度，预测能力也显然更好。

在拟合优度的视角下，GARCH 模型和隐含波动率都接受了原假设，而历史样本方差方法则拒绝了原假设，这说明了用历史样本去测算未来波动率的方法在实证上不具有可靠性，更准确的方法是使用 GARCH 模型和隐含波动率来进行估计。当然站在 $p$ 值角度再去探讨滚动估计窗与增加估计窗的优劣会发现结果仍然与我们分析的是一致的。

## 六、结论与启示

本文在风险管理在现代金融体系处于核心地位、衍生品交易全球盛行的背景下，开展了对 SPX 波动率的估计，提出了几种估计方法与检验标准。本文首先对历史波动率、隐含波动率的具体类别与理论基础进行初步介绍，历史波动率经历了估计方法由简向繁的过程，隐含波动率中则发生着一种"简化模型假设"的历史更迭。

通过实证分析，本文发现，无论从"无偏性"视角还是从"拟合优度"视角来看，隐含波动率都是几种方法中表现地最好的估计，GARCH 模型具有"过度趋势"的现象，而历史样本方差估计法虽然拥有较小的均方误差，但其波动率的稳定性实际上不符合短期段波动率的真实走势。

如果选用历史样本方差估计法来对未来波动率进行估计，那么滚动估计窗将会是更好的方法；而在 GARCH 模型中，由于过往数据仍然对趋势分析有所贡献，增加估计窗会使得预测更加准确。

从操作的简单、繁琐角度看，历史样本方差估计法具有最大的优势，因为它本质上就是计算一个无偏的样本方差，操作十分简单，GARCH 模型与隐含波动率的估计从理论上看都比较繁琐，但好在当下的统计软件都可以直接对 GARCH 模型进行参数估计与预测，且 CBOE 推出的 VIX 便是基于 Corridor 隐含波动率计算的 SPX 波动率指数，所以这两种方法也并非只有专业人士才可使用。

总而言之，无论在构建投资组合或者是风险管理的过程中，考虑资产的波动率都是非常重要的一环，作出最好的估计将有助于投资者规避风险或施行套利。

# 参考文献

# 附 录

## 附1 MATLAB 程序

```
%load data%
load X
Return=X(:,2); %Return of SP500
VIX=X(:,1); %VIX
%%%%%%%%%%%
[~,N]=size(Return); %data size
Vol=[];
%HSV & HG%
for i=1:180
    day=253+(i-1)*20; %determine the date
    RollingWindow=Return(day-252:day-1); %RollingWindow
    IncreasingWindow=Return(1:day-1); %IncreasingWindow
    Vol(i,1)=sqrt(252)*std(Return(day:day+20)); %Realized Volatility
    Vol(i,2)=sqrt(252)*std(RollingWindow); %HSV Rolling
    Vol(i,3)=sqrt(252)*std(IncreasingWindow); %HSV Increasing
    Mdl=garch(2,1);
    Mdl.Distribution='t';
    EstD1=estimate(Mdl,RollingWindow); clc
    Vol(i,4)=min(2*Vol(max(1,i-1),1),sqrt(252*forecast(EstD1,1))); %HG Rolling
    EstD2=estimate(Mdl,IncreasingWindow); clc
    Vol(i,5)=min(2*Vol(max(1,i-1),1),sqrt(252*forecast(EstD2,1))); %HG Increasing
    Vol(i,6)=VIX(day)/100; %VIX
end
%%%%%%%%%%
%MSE%
MSE=[];E=[];
for i=2:6
    E1=Vol(:,i)-Vol(:,1);
    E=[E,E1];
    MSE=mean(E.^2);
end
vvar=var(Vol(:,2:6));
p_SE=1-chi2cdf(180*MSE./vvar,180);
%%%%%
```